\documentclass[aps,prb,superscriptaddress,twocolumn]{revtex4-1}

\usepackage{graphics}
\usepackage{graphicx} 
\usepackage{dcolumn}
\usepackage{bm}
\usepackage{epstopdf}
\usepackage[usenames, dvipsnames]{color}
\usepackage{amsmath}

\begin{document}


\title{Magnetic structure and exchange interactions in the layered semiconductor CrPS$_4$}



\author{S.~Calder}
\email{caldersa@ornl.gov}
\affiliation{Neutron Scattering Division, Oak Ridge National Laboratory, Oak Ridge, Tennessee 37831, USA.}

\author{A.~V.~Haglund}
\affiliation{Department of Materials Science and Engineering, University of Tennessee, Knoxville, TN 37996.}

\author{Y.~Liu}
\affiliation{Neutron Scattering Division, Oak Ridge National Laboratory, Oak Ridge, Tennessee 37831, USA.}

\author{D.~M.~Pajerowski}
\affiliation{Neutron Scattering Division, Oak Ridge National Laboratory, Oak Ridge, Tennessee 37831, USA.}

\author{H.~B.~Cao}
\affiliation{Neutron Scattering Division, Oak Ridge National Laboratory, Oak Ridge, Tennessee 37831, USA.}

\author{T.~J.~Williams}
\affiliation{Neutron Scattering Division, Oak Ridge National Laboratory, Oak Ridge, Tennessee 37831, USA.}

\author{V.~O.~Garlea}
\affiliation{Neutron Scattering Division, Oak Ridge National Laboratory, Oak Ridge, Tennessee 37831, USA.}

\author{D.~Mandrus}
\affiliation{Department of Materials Science and Engineering, University of Tennessee, Knoxville, TN 37996.}
\affiliation{Materials Science and Technology Division, Oak Ridge National Laboratory, Oak Ridge, TN 37831.}

 

\begin{abstract}	
Compounds with two-dimensional (2D) layers of magnetic ions weakly connected by van der Waals bonding offer routes to enhance quantum behavior, stimulating both fundamental and applied interest. CrPS$_4$ is one such magnetic van der Waals material, however, it has undergone only limited investigation. Here we present a comprehensive series of neutron scattering measurements to determine the magnetic structure and exchange interactions. The observed magnetic excitations allow a high degree of constraint on the model parameters not normally associated with measurements on a powder sample. The results demonstrate the 2D nature of the magnetic interactions, while also revealing the importance of interactions along 1D chains within the layers. The subtle role of competing interactions is observed, which manifest in a non-trivial magnetic transition and a tunable magnetic structure in a small applied magnetic field through a spin-flop transition. Our results on the bulk compound provide insights that can be applied to an understanding of the behavior of reduced layer CrPS$_4$.
\end{abstract}

\maketitle

\section{Introduction}

Confining the interactions in a material to two-dimensional (2D) layers can provide routes to new phenomena by enhancing quantum behavior and increasing correlations \cite{Novoselovaac9439}. This is exemplified by broad and long-standing interest into quasi-2D compounds that host a variety of properties from unconventional superconductivity \cite{RevModPhys.60.585, doi:10.1021/ja800073m} to Kitaev physics \cite{Banerjee1055} and topologically protected Skyrmion spin structures \cite{ISI:000239792700042, Batista_2016}. The isolation of the first freestanding monolayer, graphene, and the observation of a host of remarkable properties revolutionized research into 2D materials \cite{NatureGraphene}. 2D material research has extended beyond graphene to the general class of van der Waals (vdW) materials in which the 2D layers are readily cleavable due to the weak interactions between the layers \cite{doi:10.1021/acsnano.5b05556}. These materials offer routes to new properties and behavior, with one strong focus being on 2D magnetism, which if combined with semiconductivity would open up avenues to realize spintronic devices \cite{Ando1883}.

Magnetic compounds with 2D layers that are weakly connected by vdW bonding include Cr$X$Te$_3$ ($X$=Si,Ge,Sn) \cite{Carteaux_1995, PhysRevB.92.144404, PhysRevLett.123.047203, PhysRevB.100.060402}, $M$P$X$$_3$ ($M$=Fe, Mn, Co, Ni and $X$=S, Se)\cite{PhysRevB.46.5425,Rule_2009,Wildes_2012,PhysRevB.76.134402,Wildes_2017,PhysRevLett.121.266801}, VSe$_2$ \cite{PhysRevLett.105.136805,doi:10.1021/ja207176c} and Fe$\rm _{3-x}$GeTe$_2$ \cite{FGTdoi:10.1002/ejic.200501020, PhysRevB.93.014411, ISI:000442526400011, ISI:000430389100006, PhysRevLett.122.217203,PhysRevB.99.094423, Wangeaaw8904}. In the 2D atomic layer limit for Heisenberg spins the Mermin-Wagner theorem prohibits magnetic order \cite{PhysRevLett.17.1133}, however, Onsager's theorem for Ising spins supported the stabilization of long range 2D order, which would apply to materials with strong magnetocrystalline anisotropy \cite{PhysRev.65.117}. This was shown in atomically thin 2D layers with the observation of antiferromagnetism in FePS$_3$ and ferromagnetism in CrI$_3$ and CrGeTe$_3$ \cite{doi:10.1021/acs.nanolett.6b03052, ISI:000402823400033, ISI:000402823400032}. While the reduction of the layers in a material down to atomically thin slabs typically shows a concurrent decrease in magnetic ordering temperature, this trend has been overcome to further promote routes to functionality of 2D vdW magnets. For example through the use of ionic gating in Fe$\rm _{3-x}$GeTe$_2$ to recover room temperature magnetism \cite{GateTuneableFGT} and additionally predictions of enhanced ordering temperatures in the monolayer Cr$X$Te$_3$ \cite{PhysRevB.91.235425,PhysRevB.92.035407}.

The intrinsic semiconducting magnet CrPS$_4$ belongs to this small but expanding class of 2D vdW magnets, yet has undergone only limited investigations. CrPS$_4$ is reported to be a layered antiferromagnet (AFM), however, the data presented here shows a FM model may be applicable in the atomic layered limit. While there has been a focus on ferromagnetic 2D vdW materials for applications, AFM interactions have been shown to offer a viable and alternative route to functionality \cite{RevModPhys.90.015005, ISI:000440583300013}, as well as providing a richer spin structure phase space for fundamental studies. 

In terms of AFM 2D vdW magnets the series $M$P$X$$_3$ ($M$=Fe, Mn, Co, Ni and $X$=S, Se) has formed the core of investigations. While the space groups are related, CrPS$_4$ has a distinctly different layered structure consisting of magnetic ions on a 2D rectangular lattice as opposed to hexagonal layers in $M$P$X$$_3$. CrPS$_4$, therefore, provides alternative avenues for magnetic interactions in the 2D layers not found in the $M$P$X$$_3$ series \cite{PhysRevB.91.235425}, such as the potential for interlayer anisotropic exchange interactions between neighboring magnetic ions or quasi-1D interactions. 

CrPS$_4$ has been shown to be an indirect band gap semiconductor based on optical measurements of a band gap of $\sim$1.40 eV \cite{LOUISY197861} and 1.31(1)eV \cite{doi:10.1021/acsnano.7b04679}. This is in-line with expectations from density functional theory (DFT) calculations for both functionals employed in Ref.~\onlinecite{PhysRevB.94.195307}, namely Perdew-Burke-Ernzerhof (PBE) and van der Waals density functional with optimized Becke88 parameterizations (vdW-DF-optB88). The smaller reported band gap of 0.166 eV from resistivity measurements may be a consequence of the fitting procedure  \cite{Pei_doi:10.1063/1.4940948, PhysRevB.94.195307}. The combination of magnetism, semiconductivity and 2D layers places CrPS$_4$ into a sought after category for multifunctional spintronic devices. 

The nature of the magnetism in CrPS$_4$ has not been investigated in detail with direct measurements. Different magnetic ground states have been proposed from DFT calculations \cite{PhysRevB.94.195307, Joe_2017} or from fitting to the magnetic susceptibility \cite{Pei_doi:10.1063/1.4940948}. Initial synthesis and characterization of CrPS$_4$ was first reported in Ref.~\onlinecite{Diehl:a14649} with the growth of black lustrous single crystals. The crystal structure was found to be monoclinic with the non-centrosymmetric space group $C_2$ ($\# 5$), however the structure is close to being centrosymmetric. The Cr$^{3+}$ ions are surrounded by S anions in a slightly distorted octahedral environment. This distortion could provide a route for the introduction of local anisotropy, as observed in the 2D vdW compound CrSiTe$_3$ \cite{PhysRevB.92.144404}. Shortly after the initial reports of synthesis of CrPS$_4$, magnetic susceptibility measurements observed an indication of antiferromagnetic order below 36 K \cite{LOUISY197861}, which was replicated in a recent study \cite{Pei_doi:10.1063/1.4940948}. The two-dimensional nature of the crystal structure was noted in the initial reports of CrPS$_4$ and stability of monolayers was recently predicted \cite{PhysRevB.94.195307}. 

Here, we determine the magnetic structure and exchange interactions in CrPS$_4$ through a series of neutron scattering measurements that reveal quasi-2D behavior in the bulk compound with potential for FM ordering in the atomic layered limit. We show that the previously reported transition at 36 K actually involves a subtle two-stage spin-reorientation transition. Applied-field measurements reveal the previously observed transition as a function of field and temperature is consistent with  a spin-flop transition \cite{Pei_doi:10.1063/1.4940948}. Inelastic neutron measurements (INS) provide a highly constrained exchange interaction model Hamiltonian. This is despite the measurements being on a powder sample since a balance of interaction strengths leads to a sharp separation of observed magnetic excitations due to a region of anomalously low intensity in the measured S(Q, $\omega$). Collectively, our findings show CrPS$_4$ is a quasi-2D material with ferromagnetic interactions in the layers and only weak antiferromagnetic exchange interactions between the layers, with small perturbations of field and lattice able to tune the magnetic structure. 

\section{Experimental Details}

\begin{figure}[tb]
	\centering         
	\includegraphics[trim=0cm 17cm 9cm 0cm,clip=true, width=0.85\columnwidth]{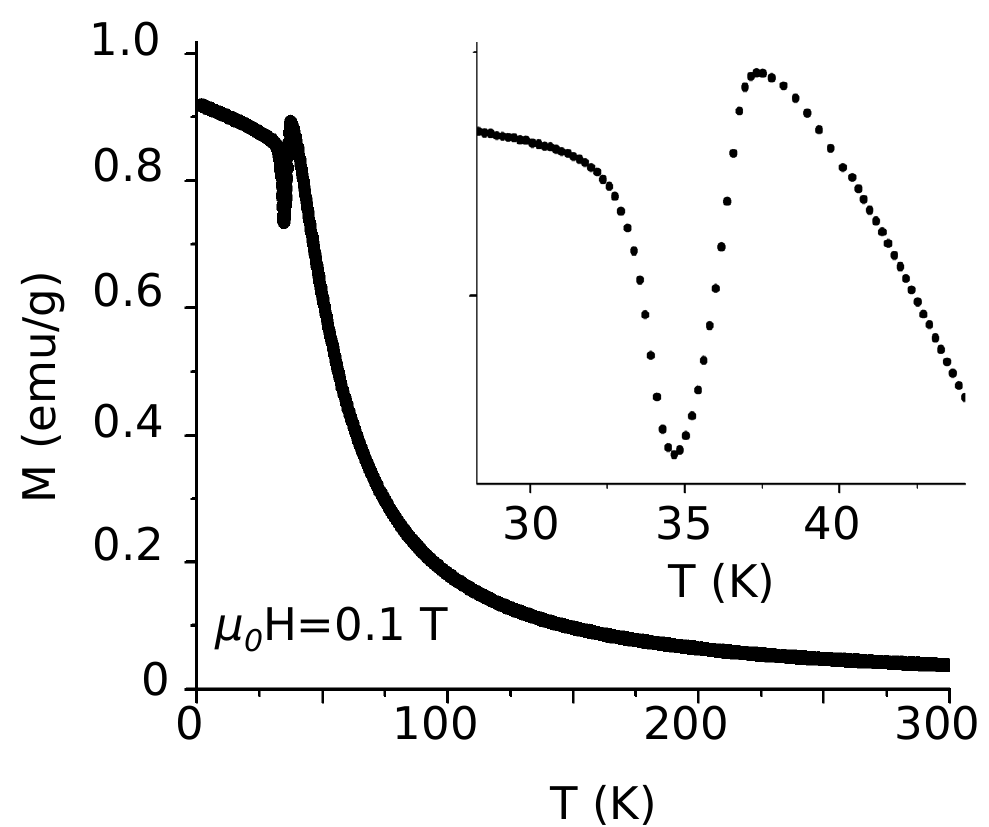}           
	\caption{\label{MagSusceptibility} Magnetic susceptibility on a powder sample of CrPS$_4$.}
\end{figure} 

Single crystals of CrPS$_4$ were grown by chemical vapor transport (CVT) following the procedure described in Ref.~\onlinecite{Pei_doi:10.1063/1.4940948}. Powders were prepared by grinding single crystals. The behavior of the single crystal and powder samples were found to be identical from magnetic susceptibility measurements and neutron scattering measurements. For all powder measurements a 5 g sample was probed in an Al can to ensure the lowest background for elastic and inelastic scattering. The sample can had a diameter of 15 mm. The single crystal measurements were performed on crystals up to 30 mg attached to a Al pin.
  
Neutron powder diffraction was carried out on the HB-2A powder diffractometer at the High Flux Isotope Reactor (HFIR), Oak Ridge National Laboratory (ORNL) \cite{Garlea2010, doi:10.1063/1.5033906}. A germanium monochromator was used to select a wavelength of 2.41 $\rm \AA$ from the Ge(113) reflection.  The pre-mono, pre-sample and pre-detector collimation was open-21'-12' and a PG filter was placed before the sample to remove higher order reflections.  The sample was loaded into a 15mm diameter Al can inside a top-loading closed cycle compressor. The choice of Al can over V can allows for a lower background set-up. No Bragg peaks are present from the Al can below 2.6 $\rm \AA^{^-1}$ and so this does not interfere with the magnetic structure analysis. The diffraction pattern was collected by scanning a 120$^{\circ}$ bank of 44 $^3$He detectors in 0.05$^{\circ}$ steps to give 2$\theta$ coverage from 5 to 130$^{\circ}$. Symmetry allowed magnetic structures were considered using both representational analysis with SARAh \cite{sarahwills} and magnetic space groups with the Bilbao Crystallographic Server \cite{Bilbao_Mag}. Rietveld refinements were performed with Fullprof \cite{Fullprof}. 

Single crystal measurements were performed on the HB-3A four-circle diffractometer at HFIR. A wavelength of 1.551 $\rm \AA$ was selected using the double focusing Si(220) monochromator \cite{Chakoumakos:ko5139}. The temperature was controlled with a bottom-loading closed cycle compressor.

Applied magnetic field measurements on a single crystal were carried out with white-beam Laue diffraction on CORELLI \cite{Ye:ut5001} at the Spallation Neutron Source (SNS) from 5 to 60 K in fields up to 5 T in a vertical field $^4$He cryomagnet. The field direction was along the $c$-axis with the crystal mounted in the horizontal (HK0) plane. To obtain a large coverage in reciprocal space the crystal was rotated in the horizontal plane over 180$^{\circ}$ in 2$^{\circ}$ steps. The correlation chopper was used to isolate the quasi-elastic only scattering from the total scattering. The data was reduced using Mantid \cite{ARNOLD2014156}.

INS measurements were performed on the time-of-flight direct geometry spectrometers HYSPEC and CNCS at the SNS, ORNL and the HB-3 triple axis spectrometer at HFIR, ORNL. The HYSPEC measurements were performed at 2 K, 36 K, 60 K and 200 K with incident energies (E$\rm _i$) of 7.5, 15, 25 and 55 meV, selected by a Fermi chopper with frequencies 240 Hz for E$\rm _i$=7.5-25 meV and 300 Hz for E$\rm _i$=55 meV, to cover the full bandwidth of magnetic excitations. The CNCS measurements were performed at 2 K with an E$\rm _i$=1.0 meV to provide an accurate measurement of any spin gap with optimal energy resolution. The HB-3 triple axis was used to perform variable temperature scans from 4 - 60 K with a fixed final energy of E$\rm _f$=14.7 meV. The collimation was 48'-40'-40'-120' with a PG(002) monochromator and analyzer. A PG filter was placed after the sample. The temperature studies on HYSPEC and CNCS utilized a top-loading $^4$He cryostat and a bottom-loading closed cycle compressor on HB-3. The data collected at the SNS was reduced using Mantid that includes vanadium normalization and corrections for time independent background \cite{ARNOLD2014156}. INS data was modeled with SpinW \cite{spinW}. The instrumental resolution was based on cuts taken of the elastic line with a FWHM resolution of 0.96 meV for E$\rm _i$=25meV, 0.6 meV for  E$\rm _i$=15 meV and 0.025 meV for E$\rm _i$=1 meV. All the INS data shown has been Bose-factor corrected. 

\begin{figure}[tb]
	\centering         
	\includegraphics[trim=0cm 0.2cm 0cm 0cm,clip=true, width=1.0\columnwidth]{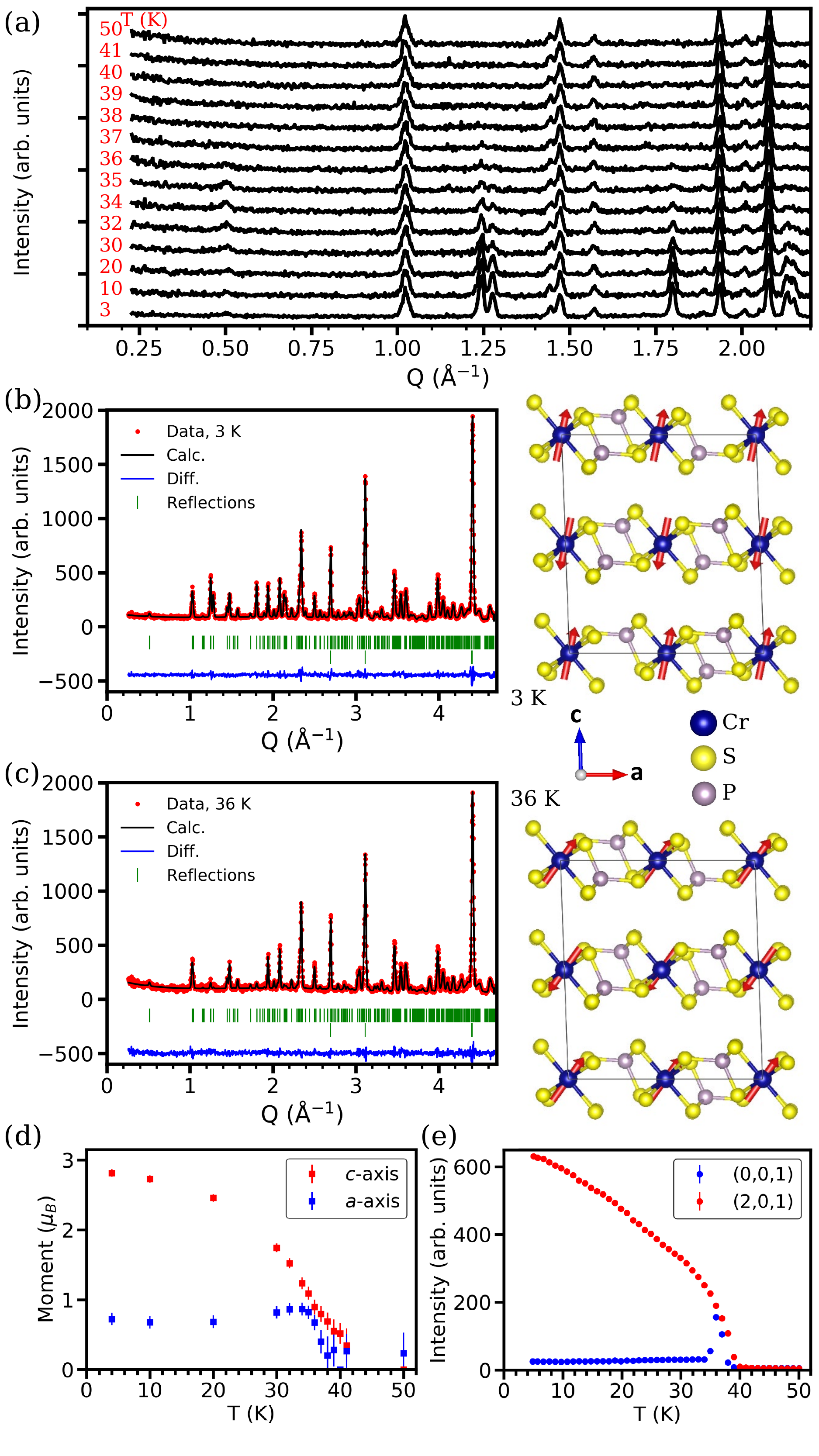}           
	\caption{\label{MagStr} (a) Neutron powder diffraction of CrPS$_4$ in the temperature range 3 - 50 K. The data has been offset in the y-axis by a constant for clarity. Refinement of the data collected and magnetic model at (b) 3 K and (c) 36 K. The magnetic and crystal structure is shown for each case. (d) Ordered Cr moment from refinement of neutron powder diffraction data. (e) Intensity of select reflections as a function of temperature from single crystal data. The magnetic unit cell is used throughout the figure.}
\end{figure} 

\begin{table}[b]
	\caption{\label{FPtableperams}Refined crystal structure parameters for CrPS$_4$ at 4 K for space group $C_2$ ($\# 5$)  with lattice constants a=10.8510(3)$\rm \AA$,  b=7.2560(2)$\rm \AA$, c=6.1058(3)$\rm \AA$, $\alpha$=90$^{\circ}$, $\beta$=92.036(4)$^{\circ}$, $\gamma$=90$^{\circ}$.}
	\begin{tabular}{c c c c c}
		\hline 
		Atom  &  $x$ & $y$& $z$ & site   \\ \hline
		Cr1  &   0 &  0.0375(39) &  0  &   $2a$ \\
		Cr2  &   0 & 0.5427(29) & 0    & $2a$ \\
		P1   &   0.2946(5) & 0.2738(42) & 0.1641(12)  &   $4c$ \\
		S1   &   0.1430(26) & -0.0019(41) & 0.6936(86)  &   $4c$ \\
		S2   &   0.1279(24) & 0.5304(40) & 0.7025(83)  &   $4c$ \\
		S3   &   0.1042(8) & 0.2845(51) & 0.1955(20) &    $4c$ \\
		S4   &   0.1271(9) & 0.7612(50) & 0.1536(18)  &    $4c$ \\
		\hline
	\end{tabular}
\end{table}

\section{Results and Discussion}

\subsection{Determining the Magnetic Structure}

Reported magnetic susceptibility measurements of CrPS$_4$ indicated the onset of magnetic order around 36 K \cite{LOUISY197861, Pei_doi:10.1063/1.4940948}, however the magnetic structure remains unresolved with conflicting predictions \cite{Pei_doi:10.1063/1.4940948, PhysRevB.94.195307}. To directly determine the magnetic structure, neutron diffraction measurements were performed on both powder and single crystal CrPS$_4$. On cooling below 40 K, additional Bragg peaks were observed at reflections distinct from those observed for the crystal unit cell, indicating long range antiferromagnetic ordering. Between 3 - 40 K the observed magnetic reflections did not change position, however, the temperature dependence of the intensity varied between reflections. This is most clearly seen by comparing the scattering at Q=0.5 $\rm \AA^{-1}$ and Q= 1.25 $\rm \AA^{-1}$ in Fig.~\ref{MagStr}(a) and the reflections in Fig.~\ref{MagStr}(e). Note the reflections are given for the magnetic unit cell throughout, where the c-axis (12.2 $\rm \AA$) is doubled compared to the non-magnetic $C_2$ space group ($c$-axis of 6.1 $\rm \AA$). The Q = 0.5 $\rm \AA^{-1}$ reflection (001) shows a maximum intensity at 36 K whereas the Q= 1.25 $\rm \AA^{-1}$ (201) reflection follows the expected order parameter-like increase in intensity. This divergent intensity behavior for reflections indicates a change in spin directions.

To uncover the underlying behavior we turn to the analysis of the neutron diffraction data. The high temperature paramagnetic region was refined within the reported crystallographic space group of $C_2$ ($\#5$), we note that nearly identical results were obtained when using the $C_2/m$ ($\#12$) space group. Applying the space group $C_2$ to the data in the magnetically ordered regime indicated no structural phase transition. Table~\ref{FPtableperams} shows the structural parameters obtained at 4 K. The additional reflections observed below 40 K could all be indexed with the k=(00$\frac{1}{2}$) propagation vector in the non-magnetic unit cell $C_2$. The Bilbao Crystallographic Server was utilized to find the maximal magnetic space groups allowed from the parent non-magnetic space group $C_2$ and propagation vector. The magnetic space group ($C_c2, \#5.16$) is the only maximally allowed space group, with two settings that yield two distinct symmetry-allowed magnetic structures. For the standard setting ({\bf a},{\bf b},{\bf 2c}; 0,0,0), spins are only non-zero along the $b$-axis, while for setting ({\bf a},{\bf b},{\bf 2c}; 0,0,$\frac{1}{2}$) spins are only present in the $a$-$c$ plane. For both cases there are two inequivalent Cr sites. For all presented models the Cr ions were fixed to have the same moment size.

Applying the models to the neutron diffraction data only produced a suitable fit to the data for the case of spins in the $a$-$c$ plane, see Figs.~\ref{MagStr}(b)-(c). At 3 K the magnetic structure is described with Cr moments primarily along the $c$-axis with a small component along the $a$-direction and a total ordered moment size of 2.8(1) $\mu_B$/Cr. This is consistent with a slight reduction from the fully ordered magnetic moment of 3 $\mu_B$ for a S=3/2 ion that was predicted by DFT that gave an ordered moment of 2.58 $\mu_B$/Cr  \cite{PhysRevB.94.195307}. The temperature evolution of the magnetic structure was followed with neutron powder diffraction at various temperatures from 3 to 50 K. The change in the component of the moment in the $a$ and $c$-axis is shown in Fig.~\ref{MagStr}(d). There is an apparent subtle spin reorientation transition between 36-40 K. This was also observed in single crystal neutron diffraction measurements that followed select reflections in the temperature range 4-40 K, Fig.~\ref{MagStr}(e). This temperature evolution of the magnetic structure is observable in the magnetic susceptibility data presented in Fig.~\ref{MagSusceptibility}. We note this behavior appears consistent with the magnetic susceptibility data in the literature \cite{LOUISY197861, Pei_doi:10.1063/1.4940948}, yet no discussion of a two stage transition has been previously presented for CrPS$_4$. The underlying driving mechanism for this behavior is an open question and motivates further detailed studies. It may be an indication of the onset of only 2D layered magnetic order prior to three dimensional order, as observed in the vdW material CrI$_3$ \cite{doi:10.1063/1.5080131}. But no signatures are observed in the current data, such as rod-like diffuse scattering indicating a regime of short range 2D ordering. Conversely it could be due to subtle structural distortions, such as a change in the local octahedral environment around the Cr ion. The $b$-axis increased from 60 K to 4 K, however, no clear anomaly was associated with the observed spin reorientation.

\subsection{Magnetic Structure in an Applied Field}

\begin{figure}[tb]
	\centering         
	\includegraphics[trim=1.5cm 5.5cm 1.5cm 0.5cm,clip=true, width=1.0\columnwidth]{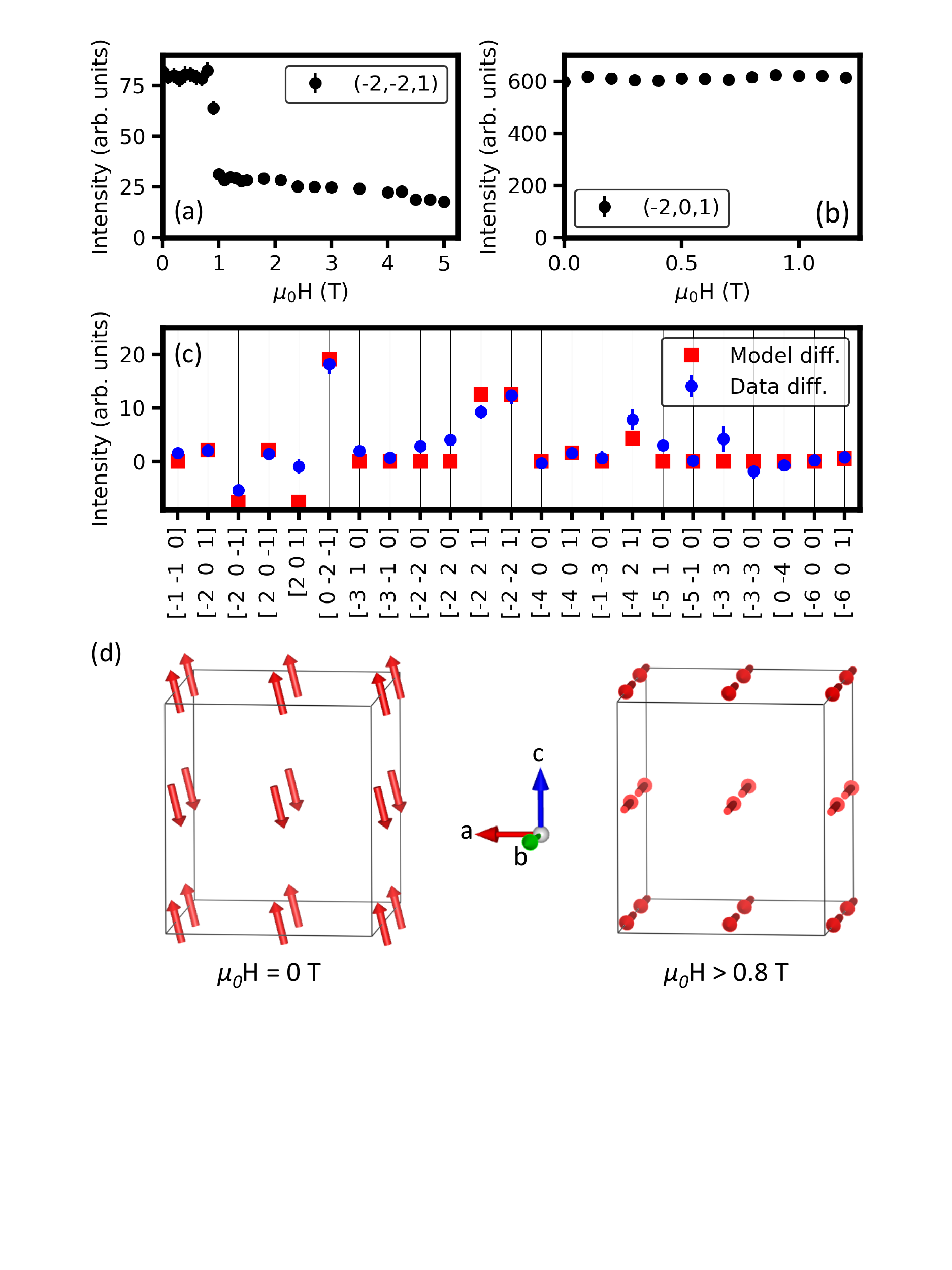}           
	\caption{\label{Fig_Magfield} Neutron diffraction measurements with an applied magnetic field parallel to the $c$-axis on a single crystal of CrPS$_4$. Field dependence at 2 K of the (a) (-2,-2, 1) and (b) (-2,0,1) reflections in the magnetic unit cell. (c) Difference of the intensity at 0 T and 1.2 T for several reflections compared to the difference in the models for $ac$-axis spin minus $b$-axis aligned moments. (d) Model spin structures in $\mu_0$H = 0 T and $\mu_0$H $>$ 0.8 T fields at 2 K.}
\end{figure} 

Having determined the magnetic structure of CrPS$_4$ we now turn to the behavior in an applied field by performing single crystal measurements. CrPS$_4$ has been reported to undergo a magnetic transition at fields under 1 T \cite{Pei_doi:10.1063/1.4940948}. The CORELLI instrument was utilized since it provides a wide coverage in reciprocal space to track changes in multiple reflections with field as well as observe any alteration in the propagation vector. For all measurements the field was applied along the $c$-axis. The measurements in 0 T and 1.2 T at 2 K showed no change in the position of the magnetic reflections with the magnetic unit cell size remaining unaltered. There was, however, changes in the intensity of certain magnetic reflections while other magnetic reflections were unaltered, see Figs.~\ref{Fig_Magfield}(a)-(b), indicating an alteration of the spin direction. The intensity of the reflections were extracted for both 0 T and 1.2 T, see Fig.~\ref{Fig_Magfield}(c). The 0 T data were modeled well with the determined low temperature magnetic structure, however no fit was possible with this $a$-$c$ spin model for the 1.2 T data. As described above there were only two maximally allowed magnetic space groups, one with spins admissible  in the $a$-$c$ plane and one with spins admissible only in the $b$-axis. The difference between the measured intensity of the reflections between 0 T and 1.2 T and the difference between the zero field model and the $b$-axis aligned model are compared in Fig.~\ref{Fig_Magfield}(c). The close qualitative agreement indicates a spin-flop transition from the zero field magnetic structure in the $a$-$c$ plane (primarily along the $c$-axis) to spins in the $b$-axis, as depicted in Fig.~\ref{Fig_Magfield}(d). We stress, however, that an unambiguous extraction of the applied field magnetic structure is beyond the limits of the data set. Further evidence for the nature of the transition can be found by inspecting the behavior presented in Ref.~\onlinecite{Pei_doi:10.1063/1.4940948} that shows the required signatures of a spin-flop transition. Namely, a sudden rotation of the spins at the critical field ($\mu_0$$\rm H_C$=0.8 T) to a direction perpendicular to the applied field and easy axis direction followed by a continuous rotation of the moments as the field in increased. Figure \ref{Fig_Magfield}(a) shows such a continued gradual change of intensity for the (-2,-2,1) reflection up to the maximum field accessible of 5 T. This may indicate a gradual canting of all spins to the applied field direction ($c$-axis), or a magnetic structure intermediate between the 0 T ($ac$-axis spins) and $>$ 0.8 T ($b$-axis) structures proposed in Fig.~\ref{Fig_Magfield}(d) that continues to rotate to $b$-axis only spins or further complex magnetic states. Within a spin-flop model this magnetic behavior occurs for systems with weak anisotropy, as opposed to a spin-flip transition which is applicable to materials with strong anisotropy. This therefore will have implications on single layer CrPS$_4$, with the size of the anisotropy being a factor in the stability.

\subsection{Magnetic Exchange Interactions}

 \begin{figure}[tb]
 	\centering         
 	\includegraphics[trim=0cm 1.9cm 5cm 0cm,clip=true, width=1.0\columnwidth]{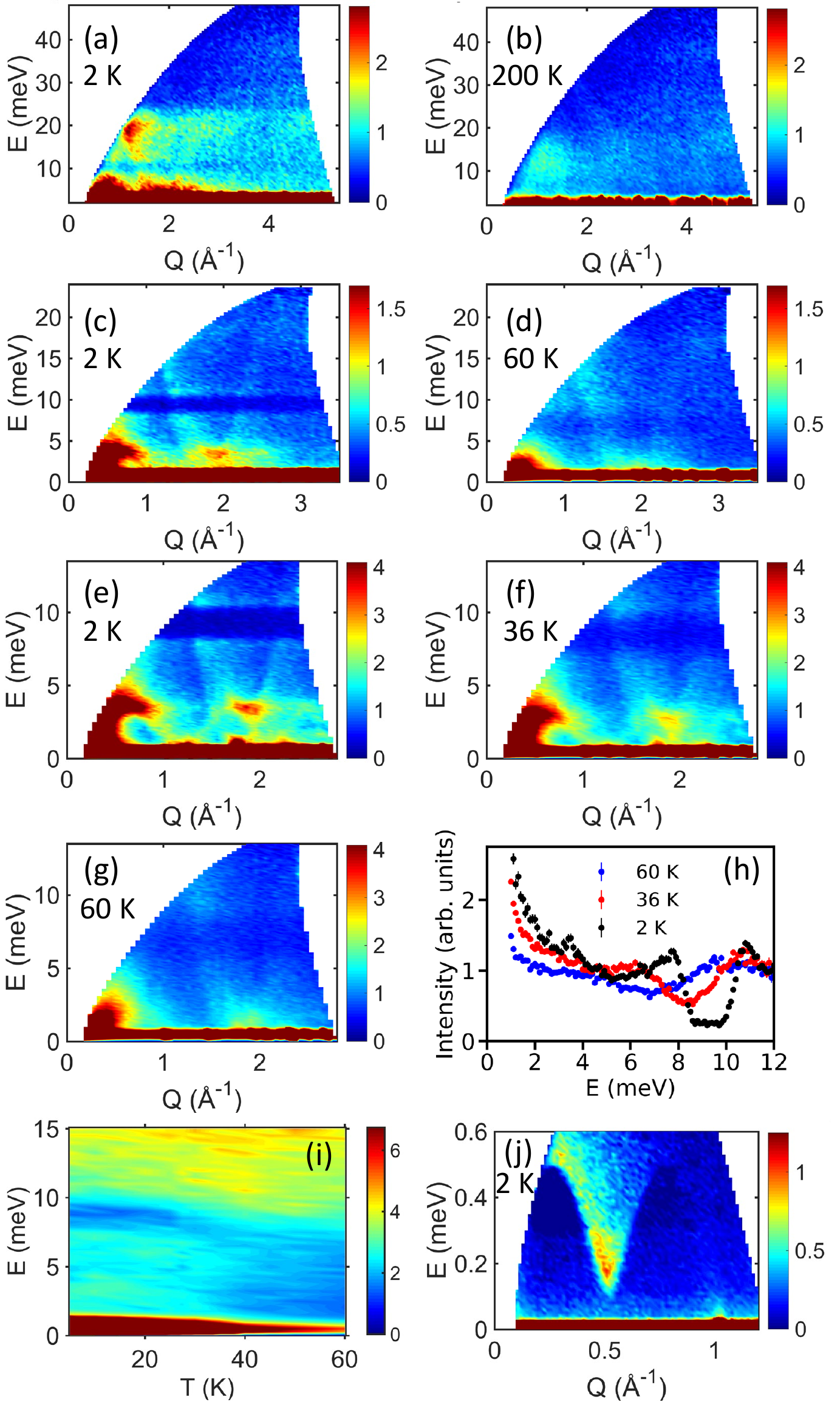}           
 	\caption{\label{INS_data} Inelastic neutron scattering  measurements on a powder sample of CrPS$_4$. Measurements on HYSPEC with an (a) E$\rm _i$=55 meV at 2 K, (b) E$\rm _i$=55 meV at 200 K, (c) E$\rm _i$=25 meV at 2 K, (d) E$\rm _i$=25 meV at 60 K, (e) E$\rm _i$=15 meV at 2 K, (f) E$\rm _i$=15 meV at 36 K and (g) E$\rm _i$=15 meV at 60 K. (h) Constant Q cuts in the range 1.25 $\le$ Q $(\rm \AA^{-1}$) $\le$ 1.5 at 2 K, 36 K and 60 K. (i) Temperature dependence of the inelastic scattering at Q=1.26 $\rm \AA^{-1}$ measured on HB-3. (j) Low energy measurements on CNCS with an E$\rm _i$=1.0 meV at 2 K.}
 \end{figure}

 \begin{figure*}[tb]
 	\centering         
 	\includegraphics[trim=0cm 1cm 1.5cm 0cm,clip=true, width=0.95\textwidth]{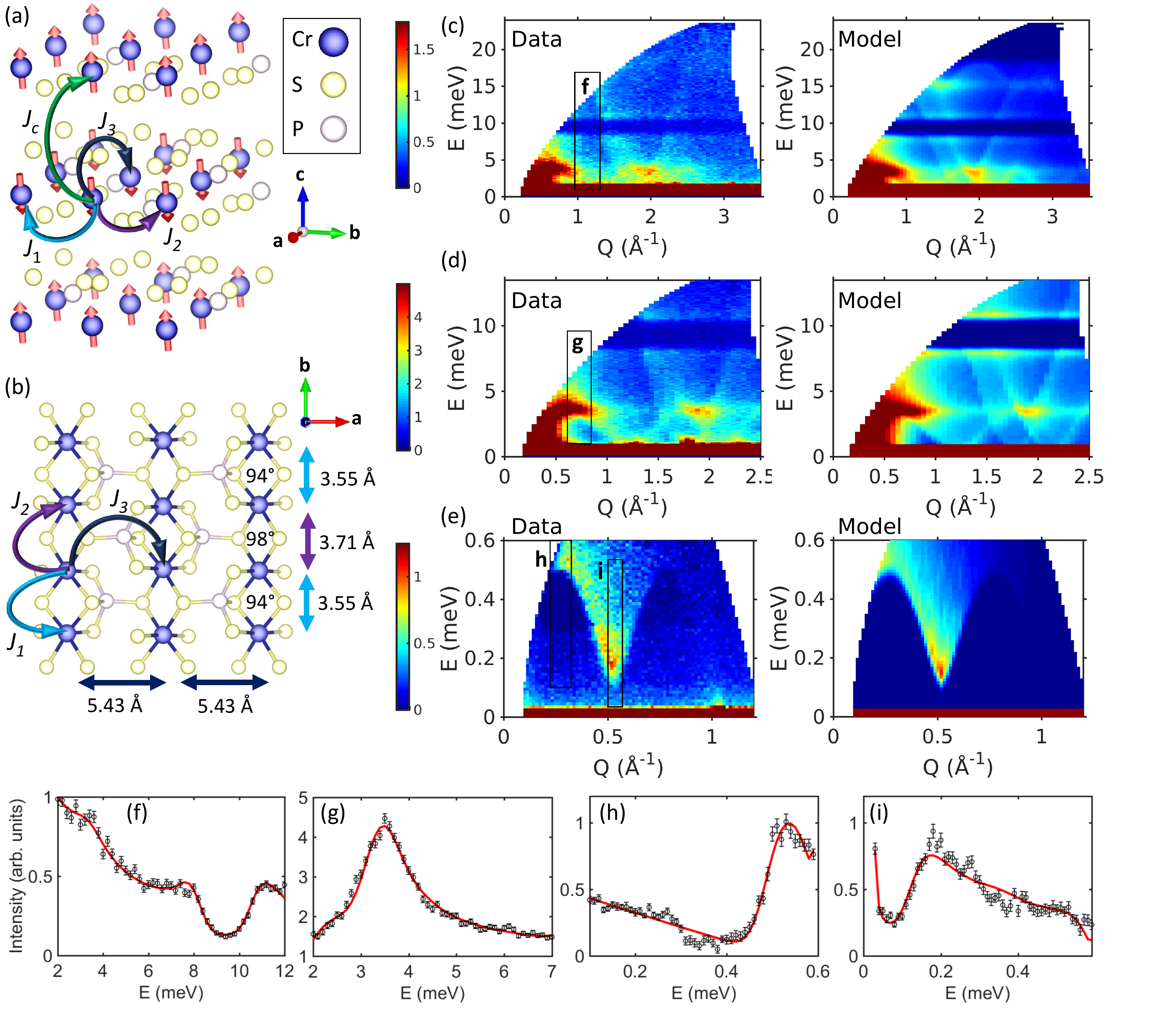}           
 	\caption{\label{INS_spinW} Determination of the exchange interactions from inelastic neutron scattering measurements. (a)-(b) Exchange interactions considered in the model. The distances and bond angles in the $ab$-plane are shown. Comparison of the 2 K data against the best fit model for different incident energies of (c) $\rm E_i$=25 meV, (d) $\rm E_i$=15 meV and (e) $\rm E_i$=1.0 meV. The elastic line has been added to the model to aid comparison. In each panel the region of 1d-cuts taken for fitting is shown by a black box and labeled g-i. The fits of the model (red line) to constant Q cuts for data (black circles) are shown, measured at (f) 1.0 $\le$ Q $(\rm \AA^{-1}$) $\le$ 1.3, $\rm E_i$=25 meV, (g) 0.7 $\le$ Q $(\rm \AA^{-1}$) $\le$ 0.9, $\rm E_i$=15 meV, (h) 0.275 $\le$ Q $(\rm \AA^{-1}$) $\le$ 0.375, $\rm E_i$=1 meV and (i) 0.5 $\le$ Q $(\rm \AA^{-1}$) $\le$ 0.55, $\rm E_i$=1.0 meV.}
 \end{figure*}

 \begin{figure}[tb]
 	\centering         
 	\includegraphics[trim=0.5cm 17.0cm 0cm 0cm,clip=true, width=1.0\columnwidth]{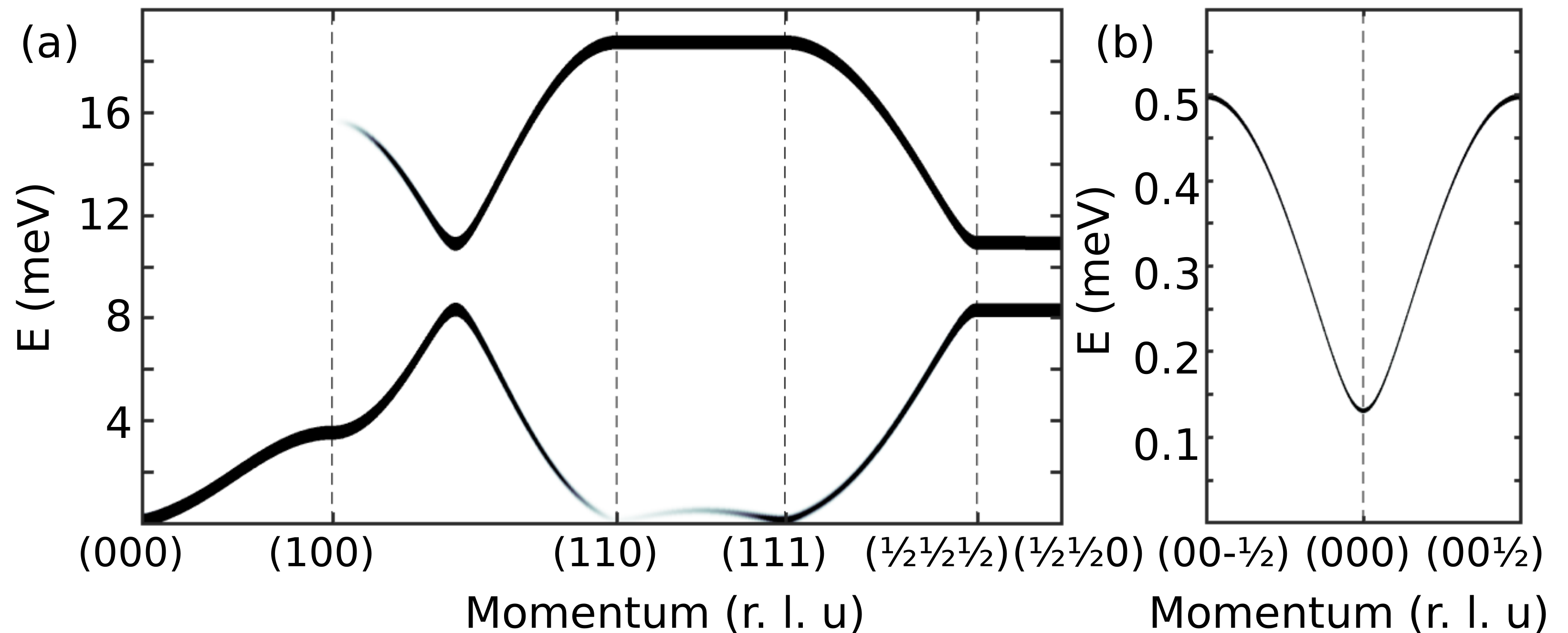}           
 	\caption{\label{INS_xtal_calc} Model of the magnon dispersion along high symmetry directions in momentum space between reciprocal lattice units (r.l.u) in the magnetic unit cell. (a) Full energy dispersion range and (b) low energy around the magnetic zone center.}
 \end{figure}

To determine the magnetic exchange interactions and anisotropy inelastic neutron scattering (INS) offers a direct and quantitative measurement. Single crystal measurements provide a full four-dimensional S(Q, $\omega$) momentum dependence in Q=HKL reciprocal space, whereas powder measurements yield a two-dimensional momentum averaged measurement of S($\rm |Q|$, $\omega$). Therefore, single crystal INS generally provides the extraction of a more robust and detailed model Hamiltonian than powder data. We note, however, that for low dimensional materials the interlayer interactions ($c$-axis for CrPS$_4$) can be much smaller than in the intralayer. This allows a more robust initial setting of parameters or even allows for the exclusion of certain interactions to a good approximation. As we show for the case of CrPS$_4$, the observed features allow the intra and inter layer exchange interactions to be defined to a high degree of accuracy, comparable to single crystal measurements, due to particularly well-defined regions in the INS data that are sensitive to distinct exchange interactions. 

The results of INS measurements are shown in Fig.~\ref{INS_data}. A high incident energy of $\rm E_i$=55 meV was utilized initially to capture the full energy bandwidth of any magnetic scattering. Figure \ref{INS_data}(a) shows inelastic features at low Q, consistent with magnetic scattering, that extend up to 20 meV at 2 K. As the temperature is increased well above the ordering temperature to 200 K inelastic excitation are still present that are consistent with quasi-elastic paramagnon scattering, Fig.~\ref{INS_data}(b). To investigate the excitations with improved resolution, further measurements were performed with incident energies of 25 meV and 15 meV, see  Figs.~\ref{INS_data}(c)-(g). A striking feature is the dispersionless near-zero intensity region centered around 9 meV. This was observed to be present for different incident energies and also on different neutron scattering instruments, confirming that it is not an instrument artifact or a spurious energy dependent feature. Moreover, tracking the energy of the feature showed a temperature dependent change through the transition as the feature shifts to lower energy and broadens. This can be seen most clearly in Figs.~\ref{INS_data}(h)-(i). Considering the low energy INS data around the elastic line with further improved resolution revealed the presence of a spin gap at magnetic Bragg positions, as observed in  Fig.~\ref{INS_data}(j). This directly shows the presence of small but finite anisotropy.

As shown above, the magnetic ground state of CrPS$_4$ has a nearly fully ordered Cr moment. This suggests that a local moment spin wave analysis of the magnetic excitations is a valid starting model to describe the observed excitations and ultimately the magnetic exchange interactions. The Hamiltonian for CrPS$_4$ is described by:

\begin{eqnarray*}
	\mathcal{H}=\sum_{i,j}J_{1} \mathbf{S}_i\cdot\mathbf{S}_j + \sum_{i,k}J_{2} \mathbf{S}_i\cdot\mathbf{S}_k + \sum_{i,l}J_{3} \mathbf{S}_i\cdot\mathbf{S}_l  \\
	+ \sum_{i,m}J_{c} \mathbf{S}_i\cdot\mathbf{S}_m +  \sum_{i,z} -D_{z} (S_i^{z})^2 
\end{eqnarray*}

where $J$ are the exchange interactions and $D$ is the single ion anisotropy (SIA) term. The full spin (S=3/2) was included in the model. The exchange interactions are shown in Figs.~\ref{INS_spinW}(a)-(b). To describe the interactions within the $a$-$b$ plane requires the introduction of three exchange interactions $J_{1}$, $J_{2}$ and $J_{3}$. A complete description of three-dimensional order requires a consideration of the exchange interaction $J_{c}$ between the layers along the $c$-axis. $J_{1}$  and $J_{2}$ on first inspection have similar distances and exchange pathways between S anions, while $J_{3}$ has a larger separation of the Cr ions with an extended pathway through two S anions. Therefore an apparent route to simplify the analysis would be to introduce the constraint  $J_{1}$=$J_{2}$ $>$ $J_{3}$. This, however, did not produce suitable models that matched the data. In particular the dispersionless low intensity region from 8-11 meV could not be reproduced. Instead, setting $J_{1}$$\ne$$J_{2}$ allowed the creation of this feature by producing separate modes of magnetic scattering from the different exchange interactions along the $b$-axis. Due to the well defined nature of this feature it was possible to highly constrain the values of $J_{1}$ and $J_{2}$ required to reproduce the energy and width of the feature. We note that the model produced identical results if $J_{1}>J_{2}$ or $J_{2}>J_{1}$, but we appeal to the well-known Goodenough-Kanamori rules in which 90$^{\circ}$ bonding favors ferromagnetism and 180$^{\circ}$ antiferromagnetism. Since the exchange interactions for $J_{1}$ are mediated by  94$^{\circ}$ bonding and $J_{2}$ by 98$^{\circ}$ bonding then $J_{1}$ should have stronger FM interactions under these rules. The low energy dispersion up to $\sim$4.5 meV was sensitive to the $J_{3}$ interaction, allowing this parameter to be well constrained. As shown in Fig.~\ref{INS_data}(j) there is an additional low energy dispersing mode below 0.5 meV. This is a result of finite inter-layer interactions and allows the $J_c$ term to be constrained. Finally, the observation of a spin gap indicated anisotropy, which was modeled with the SIA term. The origin of the SIA can be attributed to the distorted S octahedral environment around the Cr$^{3+}$ ($3d^3$) ion that leads to a small but non-zero term, as observed in related Cr$^{3+}$ 2D materials, such as CrSiTe$_3$ \cite{PhysRevB.92.144404} and CrI$_3$ \cite{Lado_2017}. This allowed an initial qualitative model to be found for exchange interactions $J_{1}$, $J_{2}$, $J_{3}$ and $J_{c}$. To go further and extract quantitative values we took cuts of the data as a function of energy over constant $|Q|$ ranges, see Figs.~\ref{INS_spinW}(f)-(i). Using suitable incident energies we were able to model the spin-gap, low energy ($\sim$4.5 meV) feature, magnetic band splitting and complete magnetic energy bandwidth in an iterative manner to find the exchange interactions that reproduced the measured S(Q, $\omega$). We note that this provided a similar quantitative modeling as would be possible with single crystal data. The final fits produced values of $J_{1}=-2.96(4)$ meV, $J_{2}=-2.09(5)$ meV, $J_{3}=-0.51(4)$ meV, $J_{c}=0.16(5)$ meV and $D_z =0.0058(5)$ meV. 

Figure \ref{INS_xtal_calc} shows plots of the momentum dependence of the magnon dispersion based on the model Hamiltonian parameters determined. The splitting of the magnon band, along with other features observed in the powder data, are observed. The energy of the spin-gap can be extracted from the model at the magnetic zone center to be 0.13(3) meV.

The analysis has focused on the magnetic excitations present in the data and was not optimzed for phonon studies. Consequently, the data collected did not show any clear evidence of anomalous phonon behavior or spin-phonon coupling that may add to the understanding of the underlying structural and magnetic phenomena.

The magnetic exchange interactions extracted for CrPS$_4$ describe spins FM-correlated in the 2D layers with the weakest interactions being between the layers. While this was expected, a further reduction in dimensionality to dominant quasi-1D chains along the $b$-axis is highlighted by the large $J_{1}$ and $J_{2}$ interactions relative to $J_{3}$. This is driven by a crystal lattice symmetry that leads to a rectangular layered motif of magnetic ions, which contrasts with the hexagonal or triangular geometries studied in most vdW 2D magnets. In addition, while we have treated the anisotropy with a SIA term, CrPS$_4$ has a non-centrosymmetric space group ($C_2$) that allows the  Dzyaloshinskii-Moriya interaction as a further term to be added to the presented Hamiltonian. Moreover, dipolar coupling was not considered. Consequently, further theoretical and experimental studies on CrPS$_4$ can offer new insights into the understanding of magnetic layered vdW materials.

\section{Conclusions}
 
Through a combination of neutron diffraction and inelastic neutron scattering, quantitative and detailed insights into the magnetic structure and exchange interactions in CrPS$_4$ were achieved. The magnetic structure has been determined to consist of ferromagnetic 2D layers antiferromagnetically coupled between the layers. This motivates efforts to achieve single layers of CrPS$_4$ with the prospect of achieving isolated ferromagnetic semiconductivity, in line with theoretical predictions \cite{PhysRevB.94.195307}. A two-stage magnetic transition was observed as a function of temperature, with an alteration of the canting of the spins in the $a$-$c$ plane. The behavior is reminiscent of that observed in the 2D layered material CrCl$_3$ where there is an initial region of ordering solely of the in-plane ferromagnetic spins prior to the onset of 3D long-range ordering \cite{doi:10.1063/1.5080131}. Additionally, it may highlight the potential sensitivity of the magnetic structure to subtle distortions that necessitates further studies. The application of a magnetic field along the easy axis drives a transition consistent with spin-flop behavior from primarily $c$-axis aligned spins to $b$-axis, offering a further tuning option to the magnetic order. The exchange interactions were robustly defined. These highlighted the 2D nature of the interactions with a small but finite interaction between the layers. The dominant in-plane interactions formed 1D chains along the $b$-axis, with the weaker interactions along the $a$-axis. The observation of a small spin-gap shows the presence of anisotropy in CrPS$_4$, which likely originates from the distorted S octahedral environment of the Cr ion. 

Collectively, the results highlight CrPS$_4$ as being a quasi-2D magnet in which the spins can be tuned by the application of small perturbations. The determination of the magnetic structure and well constrained model spin Hamiltonian presented should support further studies in the bulk down to the single layer of CrPS$_4$.     

\begin{acknowledgments}
This research used resources at the High Flux Isotope Reactor and Spallation Neutron Source, a DOE Office of Science User Facility operated by the Oak Ridge National Laboratory. DM acknowledges support from the Gordon and Betty Moore Foundation’s EPiQS Initiative, Grant GBMF9069. This manuscript has been authored by UT-Battelle, LLC under Contract No. DE-AC05-00OR22725 with the U.S. Department of Energy. The United States Government retains and the publisher, by accepting the article for publication, acknowledges that the United States Government retains a non-exclusive, paidup, irrevocable, world-wide license to publish or reproduce the published form of this manuscript, or allow others to do so, for United States Government purposes. The Department of Energy will provide public access to these results of federally sponsored research in accordance with the DOE Public Access Plan(http://energy.gov/downloads/doepublic-access-plan).
\end{acknowledgments}


%

\end{document}